\documentclass[aps,prl,showpacs,twocolumn, superscriptaddress,floatfix,a4paper]{revtex4-1}
\usepackage[utf8]{inputenc}
\usepackage{epsfig}
\usepackage[T1]{fontenc}
\usepackage{graphicx}
\usepackage{amssymb}
\usepackage{amsmath}
\usepackage{color}

\begin{document}

\title{Collective response of self-organised clusters of mechanosensitive channels}

\author{Ksenia Guseva}
\email[]{k.guseva@abdn.ac.uk}
\affiliation{Institute for Complex Systems and Mathematical Biology, King's College, University of Aberdeen, Aberdeen AB243UE, United Kingdom}
\author{Marco Thiel}
\affiliation{Institute for Complex Systems and Mathematical Biology, King's College, University of Aberdeen, Aberdeen AB243UE, United Kingdom}
\author{Ian Booth}
\affiliation{Institute of Medical Sciences, University of Aberdeen, Aberdeen AB252ZD, United Kingdom}
\author{Samantha Miller}
\affiliation{Institute of Medical Sciences, University of Aberdeen, Aberdeen AB252ZD, United Kingdom}
\author{Celso Grebogi}
\affiliation{Institute for Complex Systems and Mathematical Biology, King's College, University of Aberdeen, Aberdeen AB243UE, United Kingdom}
\author{Alessandro de Moura}
\affiliation{Institute for Complex Systems and Mathematical Biology, King's College, University of Aberdeen, Aberdeen AB243UE, United Kingdom}
\date{\today}

\begin{abstract}
  Mechanosensitive channels are ion channels activated by membrane tension. We
  investigate the influence of bacterial mechanosensitive channels spatial
  distribution on activation (gating). Based on elastic short-range interactions
  we map this physical process onto an Ising-like model, which enables us to
  predict the clustering of channels and the effects of clustering on their
  gating.  We conclude that the aggregation of channels and the consequent
  interactions among them leads to a global cooperative gating behaviour with
  potentially dramatic consequences for the cell.
\end{abstract}

\maketitle


Recent advances in the understanding of the functional organisation of the cell
membrane are shedding light on the complex dynamics of the components of the
cell surface~\cite{jacobson_revisitingfluid_1995, destainville_cluster_2008,
  wang_self-organized_2008}. There is considerable evidence that non-specific,
membrane-mediated forces are important for the formation of protein complexes on
the membrane~\cite{schmidt_cluster_2008}.  In this work, we address the
question of how the spatial organization of membrane proteins can be shaped by
their short-range interactions, and how this spatial organization can affect
their function. We focus on the behaviour of mechanosensitive channels, which
are activated by membrane deformation. These channels are present in several
organisms, such as bacteria, humans and plants, and are responsible for a
variety of functions, ranging from volume regulation, locomotion, and sensory
input and signalling~\cite{blount_mechanosensitive_2008}. Membrane mechanical
properties can also influence other types of ion
channels~\cite{perozo_physical_2002}; thus mechanosensitive channels can be used
as a general model for proteins actuated by membrane-mediated forces. The
channels which are best characterized are from E. coli, and are of two types:
channels of large (MscL) and small (MscS)
conductance~\cite{booth_mechanosensitive_2007}. These channels are responsible
for preventing the osmotic pressure from reaching dangerous levels under
hypoosmotic shock. They are activated directly by membrane tension, which causes
a membrane deformation in the channel
neighbourhood~\cite{phillips_emerging_2009}, and gate (open) when the cell is
placed in an environment of low osmolarity.  Although there are not many
mechanosensitive channels on the E. coli membrane, they are often overexpressed
for studies, and are present in larger numbers in other
organisms~\cite{blount_mechanosensitive_2008}. For these reasons, they are an
excellent and well-studied model system, from which broader conclusions may be
drawn.  Since the forces influencing them are short-range, and typically
attractive, there is the possibility that they form clusters.  Two questions of
crucial importance are: (i) under what conditions do channels cluster?  (ii) and
how does clustering affect the gating of the channels?  We address both
questions by presenting a general statistical mechanics framework ---which can
be easily carried over to other types of channels with different interaction
forces--- and conclude that channels should indeed cluster if their density (or
interaction strength) is high enough, and that this has an enormous effect on
the gating response of the channels, which display a much richer cooperative
behaviour than can be inferred from the characteristics of individual channels.

The collective behavior of MscL is a result of their mutual
interactions, mediated by the membrane deformation around them. We
model channel agglomeration and its effects on gating in a
coarse-grained manner, in which the channels are placed on a
two-dimensional lattice and only their nearest-neighbour interactions
are considered. Initially, using a lattice gas model, we obtain the
conditions for channel agglomeration, and their detailed spatial
configuration. Based on that, we subsequently show that the gating
dynamics of the clustered channels can be mapped onto an Ising-like
model, with the addition of a spatially inhomogeneous field. This
opens a new approach in the analysis of propagation of conformational
states through a cluster of proteins. One of our major findings is
that clustering leads to a lower threshold of channel activation,
causing the clustered channels to open for lower membrane tensions
than in the case of isolated channels. Furthermore, our method allows
us to study non-equilibrium properties of the system such as the
dynamics of transition. Due to this transition, clustering leads to an
increase in the time it takes for the clustered channels to open in
response to osmotic shock. Both these results show that the channel
response to osmotic stress is crucially affected by interactions
inside clusters.

\begin{figure}[ht]
  \begin{center}
    \includegraphics[width=0.98\columnwidth]{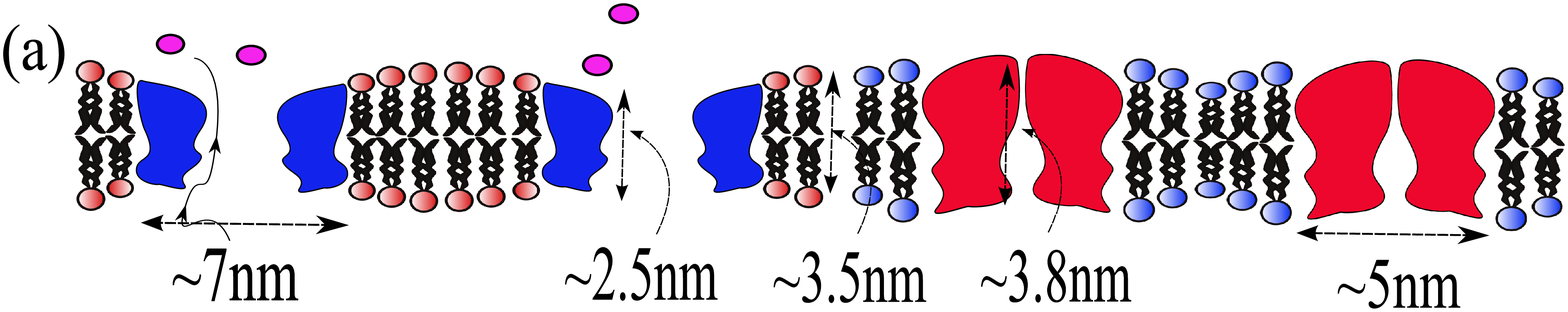}
    \hfill
    \includegraphics[width=0.82\columnwidth]{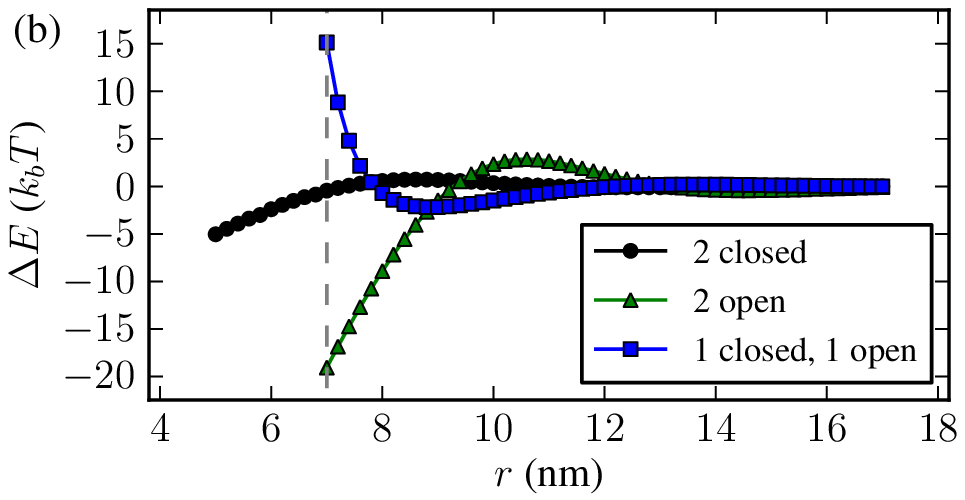}
    \vspace{-0.1cm}
  \end{center}
  \caption{(Color online) a) Examples of the deformation profile around two open
    and two closed channels respectively. b) Interaction energy between two
    channels in different states (as in~\cite{ursell_cooperative_2007}).)
\label{fig:forces}}
\end{figure}

Membrane proteins diffuse in the lipid bilayer, which can be considered as a two
dimensional fluid. They also interact with each
other~\cite{jacobson_revisitingfluid_1995}. The possible forces among membrane
proteins are electrostatic and membrane-mediated
interactions~\cite{ursell_role_2008}.  The electrostatic forces can be neglected
due to the charge screening effect in physiological solutions, since the Debye
length is $1$ nm~\cite{ destainville_cluster_2008}, which is just a fraction of
the size of a channel molecule. Segregation by lipid affinity is also not
considered since MscL does not exhibit strong lipid preferences in
\textit{E. coli}~\cite{powl_lipid-protein_2003}. Therefore, we consider only
elastic forces in what follows~\cite{ursell_role_2008}. These forces are short
range and have magnitude of the order of $\sim k_bT_0$ ($T_0$ is the typical
room temperature, $T_0 = 300 K$).  They arise from the hydrophobic mismatch
between the size of the core of a protein and the length of the lipid layer that
surrounds it~\cite{schmidt_cluster_2008}. Since lipids are more flexible than
proteins, they tend to deform and adapt to the size of the protein core (see
Fig.~\ref{fig:forces}a).  It is the energy cost of deforming the lipid layers
which results in a force among nearby channels on the membrane. As shown in
Fig.~\ref{fig:forces}a, this kind of deformation depends on whether the channel
is open or closed; therefore, the forces between channels depend on their state.
Since \textit{E. coli} is the paradigm for such studies, we use the appropriate
parameters for the MscL of these bacteria to obtain the deformation profile
$\phi(r)$, defined as the distance of the membrane contour to its relaxed state,
as a function of the distance $r$ to the protein. For that we minimise the free
energy,
\begin{equation}\label{eq:energy}
G = \int\left[\frac{K_a}{2}\left(\frac{\phi(r)}{l}\right)^2 +\frac{\kappa_b}{4}(\nabla^2 \phi(r)-c_0)^2+\tau\frac{\phi(r)}{l}\right]dr^2,
\end{equation}
where $l$ is the mismatch length, the first term represents the energy cost of
membrane stretching ($K_a = 60 k_bT_0/nm^2$), second of membrane bending
($\kappa_b = 20 k_bT_0$, with $c_0=0.009\text{nm}^{-1}$ as membrane curvature,
we considered values in the range 0--0.04 nm$^{-1}$, without any significant
difference) ~\cite{ursell_cooperative_2007,phillips_emerging_2009}, and last
term considers the membrane tension $\tau$.  To find the $\phi(r)$ which
minimises $G$, we numerically solve the Euler-Lagrange equations corresponding
to Eq. (\ref{eq:energy}) (see~\cite{ursell_cooperative_2007} for
details). Considering a system of two channels, we find that the forces between
them are short range ($\sim 5$ nm) (Fig.~\ref{fig:forces}b), and they remain
roughly the same for slight differences in value of $c_0$ and $\tau$.  This
distance is comparable to the diameter of a single protein, which validates our
coarse-grained 2D lattice approximation, which will follow.

\begin{figure}[ht]
  \centering
  \includegraphics[width=0.8\columnwidth]{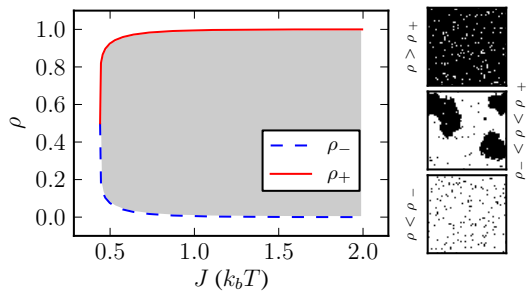}
  \caption{(Color online) Homogeneous and non-homogeneous state phase diagram,
    for $J = \frac{5k_bT_0}{4}$ and $\rho = 1.7 \times 10^{-3}$.
\label{fig:density}}
\end{figure}

\emph{The dynamics of agglomeration.} We consider initially the normal
physiological conditions for bacteria, which corresponds to a low membrane
tension. In this case, all the channels are in the closed state and diffuse on
the cell, interacting with each other through the elastic forces described
above. Using the fact that their interaction is short range, and can be
approximately decomposed into pairwise interactions, we place the channels on a
2D square lattice, and describe the system by the Hamiltonian $H = - J
\sum_{\left<ij\right>}s_is_j$, where ${s_i}$ are occupancy variables associated
with the lattice sites, which assume values either $1$ or $-1$ for an occupied
or free lattice site respectively, and $J$ is the strength of the pairwise
interaction between two channels.  The brackets in the sum indicate that it is
performed over all pairs of adjacent sites. Since the number of channels is
constant --- $\sum_is_i = N\rho$, where $N$ is the total number of sites, and
$\rho$ is the density of occupied sites --- this corresponds simply to a lattice
gas, which is characterised by the existence of either a non-homogeneous or a
homogeneous phase, depending on the interaction strength $J$ (or on the thermal
fluctuation $T$), and on the particle density $\rho$, as follows: For a weak
interaction (or a high temperature) only a homogeneous phase is observed
independently of the density.  For $J$ high enough (or $T$ low enough) this
system has two critical densities~\cite{newman_monte_1999}, $\rho_{\pm} =
\frac{1}{2}(1\pm(1-{\operatorname*{csch}}^2(2\beta J))^{\frac{1}{8}})$.  When
the density $\rho$ is lower then $\rho_{-}$ or higher then $\rho_{+}$ the
particle distribution is homogeneous. However, if $\rho_{-}<\rho<\rho_{+}$ the
particles segregate in different domains with different local densities: a low
density ($\rho_{-}$) region and a high density ($\rho_{+}$) region
(Fig.~\ref{fig:density}).  From Eq.~\ref{eq:energy}, we can determine that $J
\approx 1.25 k_bT_0$, which gives $\rho_{-} = 1.7 10^{-3}$ particles/site. Given
that an wild type \textit{E. coli} cell has on average only $5$
channels~\cite{booth_mechanosensitive_2007} and a membrane area of $\sim
6.10^{-12}$ m$^2$~\cite{phillips_physical_2008}, the density of the channels is
given by $\rho \approx 1.6\times10^{-5}$ particles/site, which corresponds to a
very low density, deep in the homogeneous phase, without any clustering. Below,
when we consider the effect of agglomeration on gating, we will argue
that there is a likely biological reason for this. Often, these channels are
artificially over-expressed, and some authors do see non-homogeneous
distributions in such
situations~\cite{norman_visualisation_2005, wahome_levels_2009},
suggesting that clustering may be occurring. Unfortunately they do not estimate
the channel density. We suggest as a possible experiment to estimate the density
in those samples and compare the appearance of agglomeration changing the
density conditions.

\begin{figure}[ht]
  \includegraphics[width=1\columnwidth]{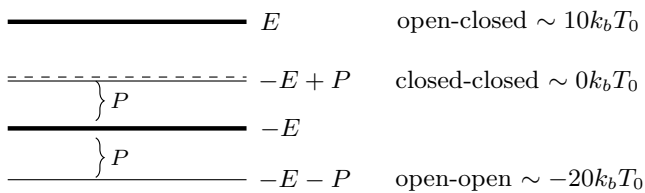}
  \caption{Interaction energies for two channels. For $E = 10 k_bT_0$
      and $P = 10 k_bT_0$
\label{fig:tabela}}
\end{figure}

\emph{Gating dynamics.} We turn now to the gating response of the channels, when
the tension is changed abruptly.  This is the case, for example, if the
osmolarity of the medium is suddenly decreased.  We note that the gating
dynamics take place on a shorter time scale than the diffusion of the channels:
the gating response of channels is of the order of
microseconds~\cite{shapovalov_gating_2004} and the free diffusion is of the
order of $\sim 0.5 nm^2/\mu s$ for crowded environment such as a biological
membrane~\cite{ramadurai_lateral_2009}. Since the area of a single channel is
approximately $\pi(2.5)^2 \sim 20 nm^2$, they cannot move significantly during a
gating event, so we assume simply that their positions remain fixed in their
initial values given by the lattice gas model discussed previously. Then the
question is: how does the spatial clustering affect the channel's response to
osmotic tension?  We will describe the state of each channel $i$ by a variable
$\sigma_i$, which can have values $1$ and $-1$, corresponding to an open or
closed state, respectively\footnote{We emphasize that the variables $\sigma_i$
  are unrelated to the variables $s_i$ used previously to describe the spatial
  configuration.}. The new energy of the system can now be written as the sum of
the non-interaction energy of the individual channels and their interaction
energy, $H = H_{\text{non}} + H_{\text{int}}$. These energies can be obtained by
solving Eq.~\ref{eq:energy}, for a system of only one or two channels,
respectively, and considering all the different channel conformation
states. This results in $H_{\text{non}} = h\sum_{i} \sigma_{i} $, where
\begin{equation}
\label{eq:h}
 h = (\Delta G_{\text{gate}} - \tau \Delta A)/2,
\end{equation}
is a global non-interaction field, where $\Delta G_{\text{gate}} \sim 50 k_bT_0$
is the energy difference between conformations, $\Delta A$ is the deformation
area of the protein, and $\tau$ is the the membrane tension, which changes
according to the osmolarity of the medium. In an analogous fashion, we can
obtain the energy levels for interacting channels, as shown in
Fig.~\ref{fig:tabela}, for an approximate distance of around 6-7 nm from their
centers. Since the interaction energies do not change significantly with
tension~\cite{ursell_cooperative_2007}, we have assumed $\tau_0 \sim 2.5
k_bT_0/nm^2$, which is the tension for which a single channel
opens~\cite{booth_mechanosensitive_2007}.
These energy levels, can be written as a Hamiltonian composed of a symmetric
term, plus an additional spatially correlated field, $H_{\text{int}} =
-\frac{E}{2} \sum_{\left<ij\right>}\sigma_i\sigma_j -
\frac{P}{4}\sum_{\left<ij\right>}(\sigma_i+\sigma_j)$. where the spatial
correlation of the second term is due to the fact that it is summed only over
\emph{neighboring} sites. The second term can be interpreted as a
non-interaction Hamiltonian, with particles subject to a local field $k_i$,
which is equal to the number of occupied neighbours of site $i$, which results
in $-\frac{P}{2}\sum_{i}k_i\sigma_i$.  Thus, the complete energy of the system
is given by the Hamiltonian
\begin{equation}
\label{hamiltonian}
 H = h\sum_{i}\sigma_i - \frac{P}{2}\sum_{i}k_i\sigma_i
 - \frac{E}{2}\sum_{\left<ij\right>}\sigma_i\sigma_j,
\end{equation}
where the sums are taken over occupied sites, and neighbouring sites.  Due to
the presence of the local field $k_i$, this is not simply an Ising Hamiltonian.
It bears resemblance to the random field Ising
model~\cite{salvat-pujol_hysteresis_2009}, where the values of $k_i$ are
replaced by randomly distributed local variables. In our case, however, they are
not random, but instead they represent a quenched \emph{correlated} disorder
which is a byproduct of the diffusion and aggregation of the channels, as
modelled by the lattice gas dynamics discussed previously.

The system given by the Hamiltonian in Eq.~\ref{hamiltonian} is investigated
using Monte Carlo simulations. We choose biologically relevant initial
conditions in which all the particles are initialised in the closed state
(mimicking conditions before an osmotic shock). This corresponds to a metastable
state for the interacting particles: The transition to a global minimum (all
channels open) involves the particles leaving a local minimum of the energy (all
channels closed), and temporarily assuming anti-aligned states with respect to
their neighbours. This process can be extremely slow, and it determines the
response time of the bacterium to osmotic shock. This is potentially an
important aspect of the system's dynamics: clustering can cause the channels'
response time to increase. We initially examine the two extreme cases where the
particle distribution is uniform. In the low-density limit ($\rho<\rho_{-}$),
most particles are isolated and there is no agglomeration. Without any mutual
iteration, the individual channels will assume a preferred state given directly
by Eq.~\ref{eq:h}, and will thus open when $\tau\Delta A > \Delta
G_\text{gate}$.  The other extreme case is when the lattice is completely
covered by particles, i.e., $\rho = 1$ and $k_i=4$ for all $i$.  In this case,
Eq.~\ref{hamiltonian} becomes the standard Ising model with external field $h -
2P$. The term $2P$ is due to the channel interactions and, as a result, the
transition to open channels occurs for a lower tension than in the case of
non-interacting particles ($P=E=0$).

\begin{figure}[ht]
\begin{center}
  \includegraphics[width=0.95\columnwidth]{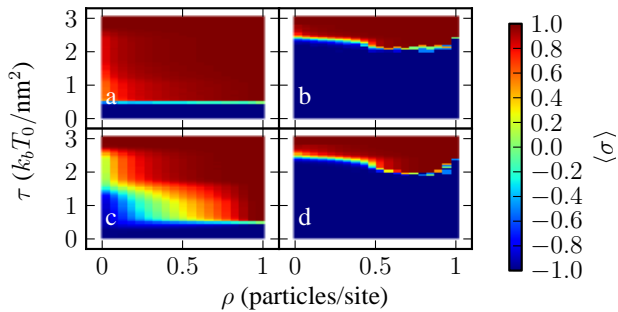}
\end{center}
  \caption{(Color online) Average conformation $\left<\sigma\right>$ as a
    function of particle density $\rho$ and membrane tension $\tau$. The top row
    corresponds to for $J = 1.25 K_bT_0$ and the bottom row to $J = 0.75
    K_bT_0$. (a) and (c) Equilibrium configuration; (b) and (d) Initial
    condition $\sigma=-1$ after a transient of $10^5$ iterations per particle;
    Simulations were carried out on a square lattice of linear size $1000$ and
    averaged over $5$ independent realizations.
    \label{fig:tension2}}
\end{figure}

For the intermediate case, when $\rho_{-}<\rho<\rho_{+}$, the process of
diffusion described previously makes the channels agglomerate in
clusters of finite size with highly irregular geometrical
structures. These structures introduce an anisotropy in the local field
$k_i$, which enables certain configurations of mixed states $-1$ and
$+1$ to coexist. This can be observed by a comparison of the gating
threshold for a group of channels inside clusters with ramified (e.g. $J
= 0.75$) or with dense (e.g. $J= 1.25$) structures, as in
Fig.~\ref{fig:tension2}. The transition between the two uniform global
states can occur in several steps, where certain discrete groups of
channels change their configuration at different values of tension. The
most obvious steps correspond to the high-density regions ($k_i=4$), and
the low-density regions ($k_i=0$), which gate for lower and for higher
tensions, respectively. However there are also intermediate transition
steps that correspond to the outer layer of the clusters (particles with
$1<k_i<4$), and the number of such steps varies with density. In the
case of compact clusters the transition for $k_i=4$ is the only one
present. If we consider now a system starting from the metastable state
where all channels are closed, after a finite (but large) simulation
time of $10^5$ Monte-Carlo steps per particle, the situation changes
significantly, as can be seen in Figs.~\ref{fig:tension2}(b) and
Figs.~\ref{fig:tension2}(d). In this case, the several discrete
transition steps are replaced by a single global transition at
significantly larger values of tension, characterizing a delay in the
reaction time of the channels.

We now turn to a more detailed comparison to the E. coli system. In bacteria the
physiological conditions correspond to a membrane tension of $\tau \sim 0.5
k_bT_0/nm^2$~\cite{booth_mechanosensitive_2007}. In this condition the
independent channels are all closed, since their gating threshold $\tau_{0}$ is
$\sim 2.5k_bT_0/nm^2$. However, our theory predicts that in this situation $\tau$ is
enough to trigger the gating response of clustered channels (see
Fig.~\ref{fig:tension2}(a)); this is a dramatic change caused by clustering on
the gating dynamics~\footnote{This decrease in the gating threshold is
  consistent with previous work done with only two
  channels~\cite{ursell_role_2007}.}.  However, because of the initial
metastable state, clustering also increases the delay in the gating response. If
long enough, this delay may be crucial, since the bacterial cell wall can only
sustain high tensions for a very limited amount of time. The gating response
time of a non-interacting channel is of the order of a few
microseconds~\cite{shapovalov_gating_2004}, and the survival time of the order
of 100 milliseconds~\cite{reutera_kinetics_2010}.  Using the approximation
derived in~\cite{cirillo_metastability_1998}, we estimate the response time for
$\rho=1$ to be $t \sim e^{\frac{2E^2}{k_bT(2P-h)}}$ Monte-Carlo steps (mcs) per
particle. Assuming that each mcs per particle correspond to the characteristic
reaction time of a single channel, i.e., $\sim 3\mu$s, this gives us a gating
response of $\sim 20$ milliseconds, for $\tau=\tau_0$. Thus we predict that
under these conditions most cells would survive the osmotic shock even with
clustered channels, but their gating time is orders of magnitude greater than
assuming isolated channels.  This is a measurable effect with current
experimental techniques, and this delay could be crucial for cells with weakened
cell walls, for example. Also, even if the channels manage to open in time, the
formation of compact channel clusters will cause problems for the closing of the
channels (for which there will also be a delay), after the osmotic stress is
removed. Clustering has therefore dramatic consequences for the non-equilibrium
dynamics of the gating.


We would like to thank A. Vieira for a careful reading of the manuscript.
This work was supported by the BBSRC SysMo and SABR under the grants BB-F00513X
and BB-G010722.

%

\begin{thebibliography}{10}%
\makeatletter
\providecommand \@ifxundefined [1]{%
 \ifx #1\undefined \expandafter \@firstoftwo
 \else \expandafter \@secondoftwo
\fi
}%
\providecommand \@ifnum [1]{%
 \ifnum #1\expandafter \@firstoftwo
 \else \expandafter \@secondoftwo
\fi
}%
\providecommand \enquote [1]{``#1''}%
\providecommand \bibnamefont  [1]{#1}%
\providecommand \bibfnamefont [1]{#1}%
\providecommand \citenamefont [1]{#1}%
\providecommand\href[0]{\@sanitize\@href}%
\providecommand\@href[1]{\endgroup\@@startlink{#1}\endgroup\@@href}%
\providecommand\@@href[1]{#1\@@endlink}%
\providecommand \@sanitize [0]{\begingroup\catcode`\&12\catcode`\#12\relax}%
\@ifxundefined \pdfoutput {\@firstoftwo}{%
 \@ifnum{\z@=\pdfoutput}{\@firstoftwo}{\@secondoftwo}%
}{%
 \providecommand\@@startlink[1]{\leavevmode}%
 \providecommand\@@endlink[0]{}%
}{%
 \providecommand\@@startlink[1]{%
  \leavevmode
  \pdfstartlink
   attr{/Border[0 0 1 ]/H/I/C[0 1 1]}%
   user{/Subtype/Link/A<</Type/Action/S/URI/URI(#1)>>}%
  \relax
 }%
 \providecommand\@@endlink[0]{\pdfendlink}%
}%
\providecommand \url  [0]{\begingroup\@sanitize \@url }%
\providecommand \@url [1]{\endgroup\@href {#1}{\urlprefix}}%
\providecommand \urlprefix [0]{URL }%
\providecommand \Eprint[0]{\href }%
\@ifxundefined \urlstyle {%
  \providecommand \doi [1]{doi:\discretionary{}{}{}#1}%
}{%
  \providecommand \doi [0]{doi:\discretionary{}{}{}\begingroup
  \urlstyle{rm}\Url }%
}%
\providecommand \doibase [0]{http://dx.doi.org/}%
\providecommand \Doi[1]{\href{\doibase#1}}%
\providecommand \bibAnnote [3]{%
  \BibitemShut{#1}%
  \begin{quotation}\noindent
    \textsc{Key:}\ #2\\\textsc{Annotation:}\ #3%
  \end{quotation}%
}%
\providecommand \bibAnnoteFile [2]{%
  \IfFileExists{#2}{\bibAnnote {#1} {#2} {\input{#2}}}{}%
}%
\providecommand \typeout [0]{\immediate \write \m@ne }%
\providecommand \selectlanguage [0]{\@gobble}%
\providecommand \bibinfo [0]{\@secondoftwo}%
\providecommand \bibfield [0]{\@secondoftwo}%
\providecommand \translation [1]{[#1]}%
\providecommand \BibitemOpen[0]{}%
\providecommand \bibitemStop [0]{}%
\providecommand \bibitemNoStop [0]{.\EOS\space}%
\providecommand \EOS [0]{\spacefactor3000\relax}%
\providecommand \BibitemShut [1]{\csname bibitem#1\endcsname}%
\bibitem{jacobson_revisitingfluid_1995}%
  \BibitemOpen
  \bibfield{author}{%
  \bibinfo {author} {\bibfnamefont{K.}~\bibnamefont{Jacobson}}, \bibinfo
  {author} {\bibfnamefont{E.~D.}\ \bibnamefont{Sheets}},\ and\ \bibinfo
  {author} {\bibfnamefont{R.}~\bibnamefont{Simson}},\ }%
  \bibfield{journal}{%
  \bibinfo {journal} {Science}\ }%
  \textbf{\bibinfo {volume} {268}},\ \bibinfo {pages} {1441} (\bibinfo {month}
  {Jun.}\ \bibinfo {year} {1995})%
  \bibAnnoteFile{NoStop}{jacobson_revisitingfluid_1995}%
\bibitem{destainville_cluster_2008}%
  \BibitemOpen
  \bibfield{author}{%
  \bibinfo {author} {\bibfnamefont{N.}~\bibnamefont{Destainville}},\ }%
  \bibfield{journal}{%
  \bibinfo {journal} {Phys. Rev. E}\ }%
  \textbf{\bibinfo {volume} {77}},\ \bibinfo {pages} {011905} (\bibinfo {month}
  {Jan.}\ \bibinfo {year} {2008})%
  \bibAnnoteFile{NoStop}{destainville_cluster_2008}%
\bibitem{wang_self-organized_2008}%
  \BibitemOpen
  \bibfield{author}{%
  \bibinfo {author} {\bibfnamefont{H.}~\bibnamefont{Wang}}, \bibinfo {author}
  {\bibfnamefont{N.~S.}\ \bibnamefont{Wingreen}},\ and\ \bibinfo {author}
  {\bibfnamefont{R.}~\bibnamefont{Mukhopadhyay}},\ }%
  \bibfield{journal}{%
  \bibinfo {journal} {Phys. Rev. Lett.}\ }%
  \textbf{\bibinfo {volume} {101}} (\bibinfo {month} {Nov.}\ \bibinfo {year}
  {2008})%
  \bibAnnoteFile{NoStop}{wang_self-organized_2008}%
\bibitem{schmidt_cluster_2008}%
  \BibitemOpen
  \bibfield{author}{%
  \bibinfo {author} {\bibfnamefont{U.}~\bibnamefont{Schmidt}}, \bibinfo
  {author} {\bibfnamefont{G.}~\bibnamefont{Guigas}},\ and\ \bibinfo {author}
  {\bibfnamefont{M.}~\bibnamefont{Weiss}},\ }%
  \bibfield{journal}{%
  \bibinfo {journal} {Phys. Rev. Lett.}\ }%
  \textbf{\bibinfo {volume} {101}},\ \bibinfo {pages} {128104} (\bibinfo {year}
  {2008})%
  \bibAnnoteFile{NoStop}{schmidt_cluster_2008}%
\bibitem{blount_mechanosensitive_2008}%
  \BibitemOpen
  \bibfield{author}{%
  \bibinfo {author} {\bibfnamefont{P.}~\bibnamefont{Blount}}, \bibinfo {author}
  {\bibfnamefont{L.}~\bibnamefont{Yuezhou}}, \bibinfo {author}
  {\bibfnamefont{P.}~\bibnamefont{Moe}},\ and\ \bibinfo {author}
  {\bibfnamefont{I.}~\bibnamefont{Iscla}},\ }%
  in\ \emph{\bibinfo {booktitle} {Mechanosensitive Ion Channels}}\ (\bibinfo
  {publisher} {Springer Netherlands},\ \bibinfo {year} {2008})%
  \bibAnnoteFile{NoStop}{blount_mechanosensitive_2008}%
\bibitem{perozo_physical_2002}%
  \BibitemOpen
  \bibfield{author}{%
  \bibinfo {author} {\bibfnamefont{E.}~\bibnamefont{Perozo}}, \bibinfo {author}
  {\bibfnamefont{A.}~\bibnamefont{Kloda}}, \bibinfo {author}
  {\bibfnamefont{D.~M.}\ \bibnamefont{Cortes}},\ and\ \bibinfo {author}
  {\bibfnamefont{B.}~\bibnamefont{Martinac}},\ }%
  \bibfield{journal}{%
  \bibinfo {journal} {Nat Struct Mol Biol}\ }%
  \textbf{\bibinfo {volume} {9}},\ \bibinfo {pages} {696} (\bibinfo {year}
  {2002})%
  \bibAnnoteFile{NoStop}{perozo_physical_2002}%
\bibitem{booth_mechanosensitive_2007}%
  \BibitemOpen
  \bibfield{author}{%
  \bibinfo {author} {\bibfnamefont{I.~R.}\ \bibnamefont{Booth}}, \bibinfo
  {author} {\bibfnamefont{M.~D.}\ \bibnamefont{Edwards}}, \bibinfo {author}
  {\bibfnamefont{S.}~\bibnamefont{Black}}, \bibinfo {author}
  {\bibfnamefont{U.}~\bibnamefont{Schumann}},\ and\ \bibinfo {author}
  {\bibfnamefont{S.}~\bibnamefont{Miller}},\ }%
  \bibfield{journal}{%
  \bibinfo {journal} {Nat. Rev. Microbiol.}\ }%
  \textbf{\bibinfo {volume} {5}},\ \bibinfo {pages} {431–440} (\bibinfo
  {year} {2007})%
  \bibAnnoteFile{NoStop}{booth_mechanosensitive_2007}%
\bibitem{phillips_emerging_2009}%
  \BibitemOpen
  \bibfield{author}{%
  \bibinfo {author} {\bibfnamefont{R.}~\bibnamefont{Phillips}}, \bibinfo
  {author} {\bibfnamefont{T.}~\bibnamefont{Ursell}}, \bibinfo {author}
  {\bibfnamefont{P.}~\bibnamefont{Wiggins}},\ and\ \bibinfo {author}
  {\bibfnamefont{P.}~\bibnamefont{Sens}},\ }%
  \bibfield{journal}{%
  \bibinfo {journal} {Nature}\ }%
  \textbf{\bibinfo {volume} {459}} (\bibinfo {year} {2009})%
  \bibAnnoteFile{NoStop}{phillips_emerging_2009}%
\bibitem{ursell_role_2008}%
  \BibitemOpen
  \bibfield{author}{%
  \bibinfo {author} {\bibfnamefont{T.}~\bibnamefont{Ursell}}, \bibinfo {author}
  {\bibfnamefont{J.}~\bibnamefont{Kondev}}, \bibinfo {author}
  {\bibfnamefont{D.}~\bibnamefont{Reeves}}, \bibinfo {author}
  {\bibfnamefont{P.~A.}\ \bibnamefont{Wiggins}},\ and\ \bibinfo {author}
  {\bibfnamefont{R.}~\bibnamefont{{RobPhillips}}},\ }%
  in\ \emph{\bibinfo {booktitle} {Mechanosensitive Ion Channels}}\ (\bibinfo
  {publisher} {Springer Netherlands},\ \bibinfo {year} {2008})%
  \bibAnnoteFile{NoStop}{ursell_role_2008}%
\bibitem{powl_lipid-protein_2003}%
  \BibitemOpen
  \bibfield{author}{%
  \bibinfo {author} {\bibfnamefont{A.~M.}\ \bibnamefont{Powl}}, \bibinfo
  {author} {\bibfnamefont{J.~M.}\ \bibnamefont{East}},\ and\ \bibinfo {author}
  {\bibfnamefont{A.~G.}\ \bibnamefont{Lee}},\ }%
  \bibfield{journal}{%
  \bibinfo {journal} {Biochemistry}\ }%
  \textbf{\bibinfo {volume} {42}} (\bibinfo {year} {2003})%
  \bibAnnoteFile{NoStop}{powl_lipid-protein_2003}%
\bibitem{ursell_cooperative_2007}%
  \BibitemOpen
  \bibfield{author}{%
  \bibinfo {author} {\bibfnamefont{T.}~\bibnamefont{Ursell}}, \bibinfo {author}
  {\bibfnamefont{K.~C.}\ \bibnamefont{Huang}}, \bibinfo {author}
  {\bibfnamefont{E.}~\bibnamefont{Peterson}},\ and\ \bibinfo {author}
  {\bibfnamefont{R.}~\bibnamefont{Phillips}},\ }%
  \bibfield{journal}{%
  \bibinfo {journal} {{PLoS} Comput. Biol.}\ }%
  \textbf{\bibinfo {volume} {3}},\ \bibinfo {pages} {e81} (\bibinfo {month}
  {May}\ \bibinfo {year} {2007})%
  \bibAnnoteFile{NoStop}{ursell_cooperative_2007}%
\bibitem{newman_monte_1999}%
  \BibitemOpen
  \bibfield{author}{%
  \bibinfo {author} {\bibfnamefont{M.~E.~J.}\ \bibnamefont{Newman}}\ and\
  \bibinfo {author} {\bibfnamefont{G.~T.}\ \bibnamefont{Barkema}},\ }%
  \emph{\bibinfo {title} {Monte Carlo Methods in Statistical Physics}}\
  (\bibinfo {publisher} {Oxford University Press},\ \bibinfo {year} {1999})%
  \bibAnnoteFile{NoStop}{newman_monte_1999}%
\bibitem{phillips_physical_2008}%
  \BibitemOpen
  \bibfield{author}{%
  \bibinfo {author} {\bibfnamefont{R.}~\bibnamefont{Phillips}}, \bibinfo
  {author} {\bibfnamefont{J.}~\bibnamefont{Kondev}},\ and\ \bibinfo {author}
  {\bibfnamefont{J.}~\bibnamefont{Theriot}},\ }%
  \emph{\bibinfo {title} {Physical Biology of the Cell}}\ (\bibinfo {publisher}
  {Garland Science},\ \bibinfo {year} {2008})%
  \bibAnnoteFile{NoStop}{phillips_physical_2008}%
\bibitem{norman_visualisation_2005}%
  \BibitemOpen
  \bibfield{author}{%
  \bibinfo {author} {\bibfnamefont{C.}~\bibnamefont{Norman}}, \bibinfo {author}
  {\bibfnamefont{Z.}~\bibnamefont{Liu}}, \bibinfo {author}
  {\bibfnamefont{P.}~\bibnamefont{Rigby}}, \bibinfo {author}
  {\bibfnamefont{A.}~\bibnamefont{Raso}}, \bibinfo {author}
  {\bibfnamefont{Y.}~\bibnamefont{Petrov}},\ and\ \bibinfo {author}
  {\bibfnamefont{B.}~\bibnamefont{Martinac}},\ }%
  \bibfield{journal}{%
  \bibinfo {journal} {European Biophysics Journal: {EBJ}}\ }%
  \textbf{\bibinfo {volume} {34}},\ \bibinfo {pages} {396} (\bibinfo {month}
  {Jul.}\ \bibinfo {year} {2005}),\ \bibinfo {note} {{PMID:} 15812637}%
  \bibAnnoteFile{NoStop}{norman_visualisation_2005}%
\bibitem{wahome_levels_2009}%
  \BibitemOpen
  \bibfield{author}{%
  \bibinfo {author} {\bibfnamefont{P.}~\bibnamefont{Wahome}}, \bibinfo {author}
  {\bibfnamefont{A.}~\bibnamefont{Cowan}}, \bibinfo {author}
  {\bibfnamefont{B.}~\bibnamefont{Setlow}},\ and\ \bibinfo {author}
  {\bibfnamefont{P.}~\bibnamefont{Setlow}},\ }%
  \bibfield{journal}{%
  \bibinfo {journal} {Archives of Microbiology}\ }%
  \textbf{\bibinfo {volume} {191}},\ \bibinfo {pages} {403} (\bibinfo {month}
  {May}\ \bibinfo {year} {2009})%
  \bibAnnoteFile{NoStop}{wahome_levels_2009}%
\bibitem{shapovalov_gating_2004}%
  \BibitemOpen
  \bibfield{author}{%
  \bibinfo {author} {\bibfnamefont{G.}~\bibnamefont{Shapovalov}}\ and\ \bibinfo
  {author} {\bibfnamefont{H.~A.}\ \bibnamefont{Lester}},\ }%
  \bibfield{journal}{%
  \bibinfo {journal} {J. Gen. Physiol}\ }%
  \textbf{\bibinfo {volume} {124}},\ \bibinfo {pages} {151} (\bibinfo {month}
  {Aug.}\ \bibinfo {year} {2004})%
  \bibAnnoteFile{NoStop}{shapovalov_gating_2004}%
\bibitem{ramadurai_lateral_2009}%
  \BibitemOpen
  \bibfield{author}{%
  \bibinfo {author} {\bibfnamefont{S.}~\bibnamefont{Ramadurai}}, \bibinfo
  {author} {\bibfnamefont{V.~K.}\ \bibnamefont{Andrea~Holt}}, \bibinfo {author}
  {\bibfnamefont{G.}~\bibnamefont{van~den Bogaart}}, \bibinfo {author}
  {\bibfnamefont{J.~A.}\ \bibnamefont{Killian}},\ and\ \bibinfo {author}
  {\bibfnamefont{B.}~\bibnamefont{Poolman}},\ }%
  \bibfield{journal}{%
  \bibinfo {journal} {J. Am. Chem. Soc.}\ }%
  \textbf{\bibinfo {volume} {131}},\ \bibinfo {pages} {12650} (\bibinfo {year}
  {2009})%
  \bibAnnoteFile{NoStop}{ramadurai_lateral_2009}%
\bibitem{Note1}%
  \BibitemOpen
  \bibinfo {note} {We emphasize that the variables $\sigma _i$ are unrelated to
  the variables $s_i$ used previously to describe the spatial configuration.}%
  \bibAnnoteFile{Stop}{Note1}%
\bibitem{salvat-pujol_hysteresis_2009}%
  \BibitemOpen
  \bibfield{author}{%
  \bibinfo {author} {\bibfnamefont{F.}~\bibnamefont{{Salvat-Pujol}}}, \bibinfo
  {author} {\bibfnamefont{E.}~\bibnamefont{Vives}},\ and\ \bibinfo {author}
  {\bibfnamefont{M.}~\bibnamefont{Rosinberg}},\ }%
  \bibfield{journal}{%
  \bibinfo {journal} {Phys. Rev. E}\ }%
  \textbf{\bibinfo {volume} {79}},\ \bibinfo {pages} {061116} (\bibinfo {month}
  {Jun.}\ \bibinfo {year} {2009})%
  \bibAnnoteFile{NoStop}{salvat-pujol_hysteresis_2009}%
\bibitem{Note2}%
  \BibitemOpen
  \bibinfo {note} {This decrease in the gating threshold is consistent with
  previous work done with only two channels~\cite {ursell_role_2008}.}%
  \bibAnnoteFile{Stop}{Note2}%
\bibitem[{\citenamefont{Reutera et~al.}(2010)\citenamefont{Reutera, Haywardb,
  Millera, Drydenb, and Booth}}]{reutera_kinetics_2010}
\bibinfo{author}{\bibfnamefont{M.}~\bibnamefont{Reutera}},
  \bibinfo{author}{\bibfnamefont{N.~J.} \bibnamefont{Haywardb}},
  \bibinfo{author}{\bibfnamefont{S.}~\bibnamefont{Millera}},
  \bibinfo{author}{\bibfnamefont{D.~T.} \bibnamefont{Drydenb}},
  \bibnamefont{and} \bibinfo{author}{\bibfnamefont{I.~R.} \bibnamefont{Booth}},
  \bibinfo{journal}{to be published}  (\bibinfo{year}{2010}).
\bibitem{cirillo_metastability_1998}%
  \BibitemOpen
  \bibfield{author}{%
  \bibinfo {author} {\bibfnamefont{E.}~\bibnamefont{Cirillo}}\ and\ \bibinfo
  {author} {\bibfnamefont{J.}~\bibnamefont{Lebowitz}},\ }%
  \bibfield{journal}{%
  \bibinfo {journal} {J. of Stat. Phys.}\ }%
  \textbf{\bibinfo {volume} {90}},\ \bibinfo {pages} {211} (\bibinfo {month}
  {Jan.}\ \bibinfo {year} {1998})%
  \bibAnnoteFile{NoStop}{cirillo_metastability_1998}%
\end{thebibliography}
\end{document}